\documentclass{article}
\usepackage{graphicx}

\def\beq{\begin{equation}}
\def\eeq{\end{equation}}
\def\f{\frac}

\title{Maximally entangled states in the Hydrogen molecule: The role
of spin and correlation}

\author{M. Babamoradi${}^1$ \footnote{Electronic address:
babamoradi@physics.sharif.edu}, M. Heidari Saani${}^2$ and M. A.
Vesaghi${}^{1,3}$.}

\begin{document}
\maketitle

\begin{center}
$^1$Department of Physics, Sharif University of Technology, P.O.Box:
11365-9161, Tehran, Iran.\\
$^2$Physics Group, Malek Ashtar University, P.O.Box: 83145-115,
Shahin Shahr, Iran.\\
$^3$School of Physics, Institute for Studies in Theoretical Physics
and Mathematics, P.O.Box: 19395-5531, Tehran, Iran.\\
\end{center}

\begin{abstract}
By going beyond Hubbard Hamiltonian we reflected correlation effects
accurately in the wavefunctions of $H_2$. Using $ab$ $initio$ e-e
interaction parameters resulted maximally entangled ground and third
excited states. We assigned this maximally entangled character to
the nonmagnetic (S=0) property of these states and also the
minimally entangled character of the first excited states to its
magnetic property (S$\neq0$). By switching on a magnetic field an
entangled state with $S_{z}=0$ can be extracted from a minimally
entangled degenerate magnetic state. We suggest that presence of a
moderate correlated system and a non-magnetic $(S_{z}=0)$ electronic
state can be two criteria for finding maximally entangled electronic
states in a realistic molecular  system.
\end{abstract}

\section{Introduction}

Since Einstein, Podolsky and Rosen \cite{ein} and Schr\"{o}dinger
\cite{scr} investigated the non-classical properties of quantum
systems and entered new concept as entanglement in quantum physics,
it had become strange property in interaction between particles.
Recently the study of the entanglement is a useful resource for
quantum communications and information processing \cite{mac} such as
quantum teleportation \cite{bnn,bou}, superdense coding \cite{ben},
quantum key distribution \cite{eke}, and quantum cryptography
\cite{fuc} whose input states are constructed to be maximally
entangled. Also entanglement has been suggested as a quantitative
measure for electron-electron (e-e) correlation in many body systems
\cite{hua,mai}. As a simple illustration of entanglement we can say
that, if there is no way to write the states of two particles as a
product of the two systems states in the Hilbert space, then there
will be an entangled system \cite{ysl}. A lot of investigation has
been done about measuring entanglement in the fermionic systems,
such as the Wooters' measure \cite{woo} and the Schliemann's measure
\cite{scl,scm}. Through Gittings' investigation \cite{jrt}, it is
shown that all these entanglement measures are not suitable but the
Zanardi's measure \cite{zan,zar} satisfies all desirable properties
of entanglement measurement. \\ The $H_2$ molecule is the simplest
two electron systems that can be used to implement a robust many
body calculation based on Hubbard model \cite{alv}. Traditional
Hubbard model which is a priory many-body approach usually is used
as a first attempt to calculate entanglement. This model gives
maximum entanglement by setting e-e interaction parameter $U=0$,
which is a controversial conclusion \cite{zar,wan,sjg}. In this
paper we go beyond Hubbard model to calculate entanglement of the
non-magnetic ground and magnetic excited states of $H_2$ molecule.
The $H_{2}$ molecule is the simplest realistic two electron
correlated system in nature which our model can be implemented and
considering all direct and exchange interaction terms beyond Hubbard
model let us account correlation effects more accurately in the
wavefunctions. The Zanardi measurement for calculating entanglement
was employed. We will investigate the ground and excited state of
$H_2$ to find the effect of the spin on the entanglement of states
and results give maximally entangled states with non-zero U value.
We also discuss the difference between maximally, moderately and
zero entangled states based on their magnetic property and
correlation dependence.

\section{Calculation method}
Now we explain our calculation method. The complete Hubbard
Hamiltonian is defined as \cite{hei}: \beq
H=\sum_{ij\sigma}t_{ij}c^\dag_{i\sigma}c_{j\sigma}+\f{1}{2}\sum_{ijlm\sigma\sigma'}
V_{ijlm}c^\dag_{i\sigma}c^\dag_{j\sigma'}c_{m\sigma'}c_{l\sigma}.
\eeq The first term contains non-interacting part of Hamiltonian
which can be written as:

\beq \sum_{ij\sigma}t_{ij}c^\dag_{i\sigma}c_{j\sigma}=
\varepsilon_\circ\sum_{i=1,2;\sigma}c^\dag_{i\sigma}c_{i\sigma}+t\sum_{i\neq
j;\sigma}c^\dag_{i\sigma}c_{j\sigma}\eeq

$$t_{21}=t_{12}=t$$
where $\varepsilon_\circ$  is the energy of atomic orbital,
$c^\dagger_{i\sigma}$ and $c_{i\sigma}$ are fermionic creation and
annihilation operators on site \emph{i} with spin $\sigma$,
respectively and \emph{t} stands for the hopping integral between
two \emph{H} atomic sites of the electrons with the same $\sigma$.
The second term of Hamiltonian that contains e-e interaction part
can also be written as:

\begin{eqnarray} \f{1}{2}\sum_{ijlm\sigma\sigma'}
V_{ijlm}c^\dag_{i\sigma}c^\dag_{j\sigma'}c_{m\sigma'}c_{l\sigma}&=&U\sum_{i=1,2}n_{i\uparrow}n_{i\downarrow}+
\f{1}{2}J\sum_{i\neq j;\sigma\sigma'}n_{i\sigma}n_{j\sigma'}\cr
+\f{1}{2}X_1\sum_{ijlm\sigma\sigma'}c^\dag_{i\sigma}c^\dag_{j\sigma'}c_{m\sigma'}c_{l\sigma}&+&\f{1}{2}X_2\sum_{ijlm\sigma\sigma'}c^\dag_{i\sigma}c^\dag_{j\sigma'}c_{m\sigma'}c_{l\sigma}
\end{eqnarray}
 where $V_{ijlm}$ contains all coulomb interaction between electrons which
involves \emph{U} as on-site Coulomb repulsion, \emph{J} as
inter-site Coulomb repulsion and
$n_{i\sigma}=c^\dag_{i\sigma}c_{i\sigma}$ is density operator. The
last terms $X_{1}$ and $X_{2}$  are the exchange interactions
parameters that can be interpreted by quantum mechanics. We consider
two electrons in two sites or $1s$ orbital of $H_2$ molecule with
spin up and down, therefore we have four states for a single
electron so for two electrons there are $C(4,2)=6$ states which are
represented with notation
$|\varphi>=|n_{1\uparrow}n_{1\downarrow}n_{2\uparrow}n_{2\downarrow}>$
as \begin{eqnarray}
|\varphi_1>&=&|1100>,|\varphi_2>=|1010>,|\varphi_3>=|1001> \cr
|\varphi_4>&=&|0110>,|\varphi_5>=|0101>,|\varphi_6>=|0011>
\end{eqnarray}
With these sets of states, we can write Hamiltonian as:
$$H=\left(%
\begin{array}{cccccc}
  U & 0 & t+X_1 & -t-X_1 & 0 & X_2 \\
  0 & J-X_2 & 0 & 0 & 0 & 0 \\
  t+X_1 & 0 & J & -X_2 & 0 & t+X_1 \\
  -t-X_1 & 0 & -X_2 & J & 0 & -t-X_1 \\
  0 & 0 & 0 & 0 & J-X_2 & 0 \\
  X_2 & 0 & t+X_1 & -t-X_1 & 0 & U \\
\end{array}%
\right).$$
 All the diagonal elements contain a term $t_{ii}$, where
it is roughly four times of the energy of an electron in the $1s$
state of atomic hydrogen. Using $ab$ $initio$ energies for $H_2$ and
$H_2^+ $ from Ref. \cite{chi} and with the parametric energies of
Hamiltonian, parameters $\varepsilon_\circ$, \emph{t}, \emph{U},
\emph{J}, $X_1$ and $X_2$, were evaluated and are given in Table I.
\begin{table}[h]
\caption{The calculated values of Hamiltonian parameters.}
\begin{tabular}{|c|c|c|c|c|c|c|}
\hline

Hamiltonian parameter & $\varepsilon_\circ$ & $t$ & $U$ & $J$ & $X_1$ & $X_2$ \\
\hline
calculated value(eV) & -28.56 & 6.36 & 19.68 & 17.90 & 0.95 & 1.36 \\
\hline
\end{tabular}
\end{table}
\\Entanglement measurement is defined by von Neumann's entropy
as\cite{zan}: \beq S(\rho_A)=-tr(\rho_A\log_2\rho_A)\eeq where $A$
and $B$ are bipartite subsystem which in our model are $1s$ orbital
of each Hydrogen atoms in our model. $\rho_A$ is reduced density
matrix that is defined by:
$$\rho_A=tr_B\rho_0=\sum_j<j|_B(|\psi><\psi|)|j>_B$$ where
$tr_B$ stands for tracing over all sites except the $B$ sites and
$\mid j>_B$ is eigenstate for $B$ part
$(|n_{2\uparrow}n_{2\downarrow}>)$ i.e. $\mid00> , \mid01> , \mid10>
and \mid11>$. After these calculation the reduced density matrix for
the ground state (not normalized) becomes:
\beq \rho_A=tr_A|\psi\rangle\langle\psi|=\left(%
\begin{array}{cccc}
  1 & 0 & 0 & 0 \\
  0 & \alpha_+^2 & 0 & 0 \\
  0 & 0 & \alpha_+^2 & 0 \\
  0 & 0 & 0 & 1 \\
\end{array}%
\right).\eeq For other states, reduced density matrices have been
evaluated accordingly and their entanglements were calculated. We
listed the resultant entanglement values in Table II.
\begin{table}
\caption{The parametric energies of hydrogen molecule are listed in
the first column of the Table, where $
\alpha_{\pm}(x)=x\pm\sqrt{1+x^2}$ and $x=\frac{(U-J)}{4(t+X_1)}$.
The values of energies from Ref.
 \cite{chi} are listed in the second column, eigenfunction for these
states are also listed in the third column. The $S^2$ and $S_z$ of
each state are given in the next columns and the last column is the
calculated entanglement $S(\rho_A)$ for these states, based on
Zanardi's measurement.}
\begin{tabular}{|c|c|c|c|c|c|}
  \hline
  E& E(eV) & $\psi$ & $S^2$ & $S_z$ & $S(\rho_A)$ \\
  \hline
 $E_0=2\varepsilon_\circ+X_2+J+2(t+X_1)\alpha_-(x)$ & -51.60 &$ \varphi_1+\varphi_6+\alpha_+(x)(\varphi_4-\varphi_3)$ & 0 & 0 &$ \sim2$ \\
 $E_1=2\varepsilon_\circ+J-X_2 $& -40.58 & $\varphi_3+\varphi_4 $& 2 & 0 & 1 \\
 $E_1=2\varepsilon_\circ+J-X_2 $& -40.58 & $\varphi_2$ & 2 & 1 & 0 \\
 $E_1=2\varepsilon_\circ+J-X_2 $& -40.58 & $\varphi_5$ & 2 & -1 & 0 \\
 $E_2=2\varepsilon_\circ+U-X_2 $& -38.80 & $\varphi_6-\varphi_1$ & 0 & 0 & 1 \\
 $E_3=2\varepsilon_\circ+X_2+J+2(t+X_1)\alpha_+(x)$ & -22.32 & $\varphi_1+\varphi_6+\alpha_-(x)(\varphi_4-\varphi_3)$ & 0 & 0 & $\sim2 $\\
\hline
\end{tabular}
\end{table}
By acting $S^{2}$ on the $|E_0\rangle,...,\;{\rm and}\;|E_3\rangle$
eigenstates, we could find total spin of each state in which
$|E_0\rangle,\;|E_2\rangle\;{\rm and}\;|E_3\rangle$ have $S^2=0$,
but $|E_1\rangle$ has $S^2=2$.  We find eigenvalues and eigenstates
(not normalized) of hydrogen molecule as shown in Table II, where $
\alpha_{\pm}(x)=x\pm\sqrt{1+x^2}$ and $x=\frac{(U-J)}{4(t+X_1)}$.
Also the related $S_z$ values of each eigenstates have been
summarized in Table II. The eigenstates results schematically
demonstrated in Fig. 1. Now let us discuss calculation results
summarized in Table II and Fig.1.

\begin{figure}
  \includegraphics[width=10cm]{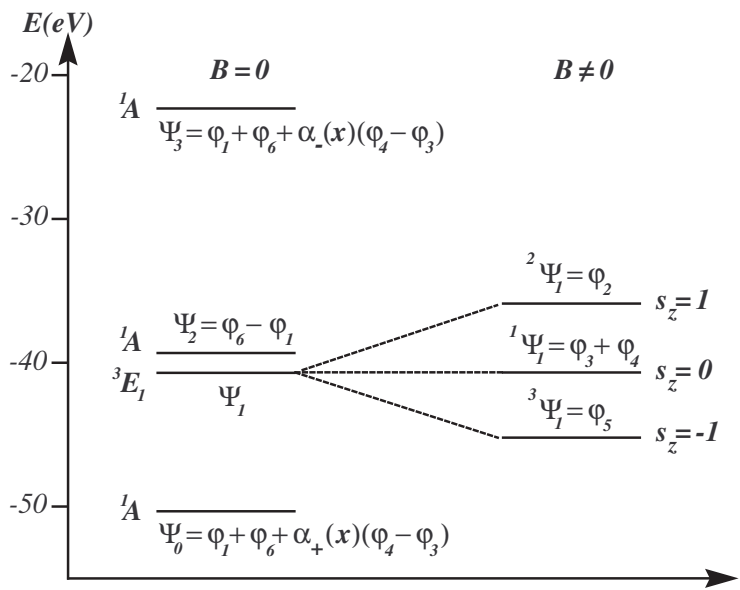}\\
  \caption{The ground and excited states for $H_2$ molecule before
and after switching on a magnetic field with their related
eigenfunctions and spins.}
\end{figure}

\section{Results and discussion}
From this table we conclude that the many electron wavefunctions
have very weaker dependence on the interaction parameters than the
energy levels. The wavefunctions of the first and second excited
states are explicitly independent from Hamiltonian parameters. The
wavefunction of the ground and third excited state are dependent
upon $e-e$ correlation parameters via
$\alpha_{\pm}=x\pm\sqrt{1+x^2}$ with $x=\frac{(U-J)}{4(t+X_1)}$ in
Table II where the exchange interaction $X_2$ is absent in the
wavefunctions. Both of these states are nonmagnetic (S=0). Using
parameters of Table I and von Neumann's entropy of Eq. (6) an
entanglement value $S[\rho]\sim2$ was obtained for these states. The
maximum available value of entanglement is $log_2 d$ for a system
with the Hilbert space dimension of the smaller subsystem as $d$
\cite{cben}. Accordingly, for $H_2$ molecule the maximum available
entanglement is 2 therefore the resultant maximally entangled ground
and third excited states in Table II can be explained by the
corresponding wave functions of these states. The wavefunctions of
the ground and third excited states are superposition of four  body
basis of the systems $\varphi_1,\;\;\varphi_3,
\;\;\varphi_4\;\;{\rm and}\;\; \varphi_6$ with equal
coefficients since the value of $\alpha_{\pm}$ from table I becomes
1.\\From Table II, the first excited state $E_1$ is a spin triplet
state with $S=1$ and its wavefunction is independent fromcorrelation
parameters. The value of entanglement for the $S_z=0$ eignfunction
is 1 where its value for the eignfunctions with $S_z=\pm1$ is zero.
The difference between entanglements of the degenerate wavefunctions
with different $S_z$ can be explained by their related
wavefunctions. The entanglement of the eignfunction of
$|E_1,S_z=0\rangle$ state is a linear combination of $\varphi_3$ and
$\varphi_4$ whereas the wave functions of other
$|E_1,S_z=\pm1\rangle$ are separable ($\varphi_2$ or $\varphi_5$).
The importance of the nonzero spin of the first excited state is
that this state can be detected by Electron paramagnetic resonance
(EPR) under some condition \cite{wyk}. When the magnetic field is
off the wavefunction is a superposition of the degenerate
wavefunctions with different $S_z$ values. We calculated
entanglement of this degenerate wavefunction as 0. However, after
switching on the magnetic field, a wavefunction with definite value
of entanglement emerges. The values of the $S_z$ and entanglement
for this state are 0 and 1 according to Table II. Therefore we
conclude that in practice by switching on a magnetic field on a
magnetic state, one can switch from a degenerate and not entangled
wavefunction to an entangled wavefunction with $S_z=0$. \\ The
second excited state $E_2$ is nonmagnetic, $S=0$, where the
wavefunction is independent from correlation parameters. The
calculated entanglement is 1. The wavefunction of this state is only
a linear combination of two basis set $\varphi_6$ and $\varphi_1$,
hence its entanglement is smaller than ground and third excited
states. By comparing the values of the entanglement listed in Table
II, we conclude that the nonmagnetic states $(S=0)$ in which the
wavefunction depend on interaction parameters are maximally
entangled and the magnetic states whose wavefunctions are
independent from correlation parameters are not entangled.\\Using
von neumann's entropy in Eq. 6, we calculated the variation of
entanglement for maximally entangled ground state of $H_2$ molecule
with respect to the combination of correlation parameters
$x=\frac{(U-J)}{4(t+X_1)}$. Results were plotted in Fig. 2.

\begin{figure}
  \includegraphics[width=25pc]{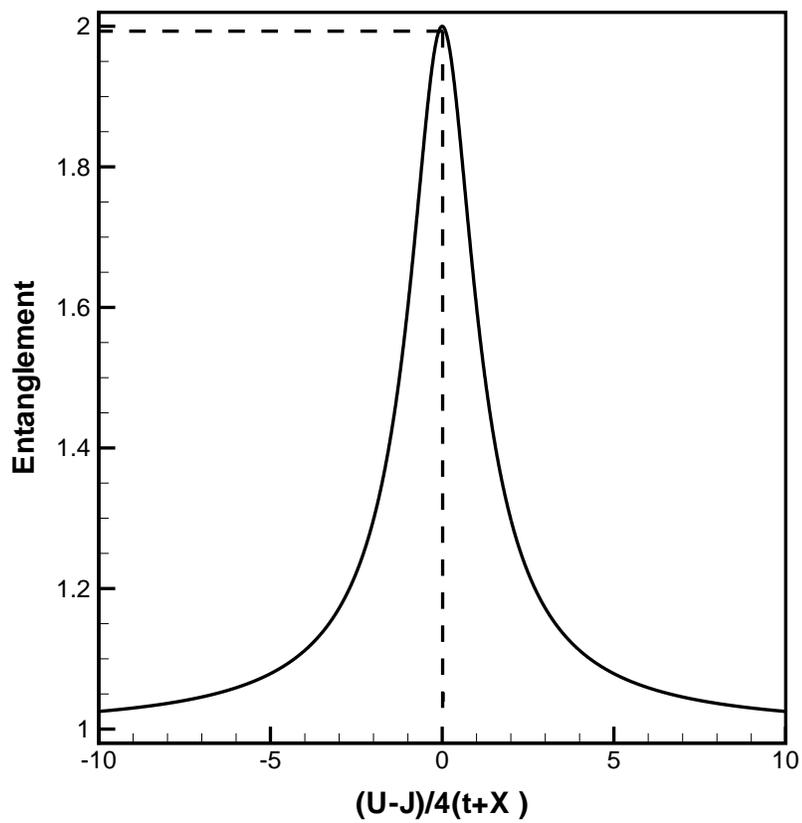}\\
  \caption{Entanglement for the ground state of Hydrogen molecule
versus $e-e$ interaction parameters.}
\end{figure}

 In this figure we observe that the value of entanglement is maximum with $x=0$
($U=J$). This conclusion resolves previously reported results from
other groups \cite{zar,wan,sjg} who obtained maximally entangled
ground state with  $U=0$. The  $U=0$ result imply that the maximally
entangled ground state is not attainable for the $H_2$ since the
$ab$ $initio$ value of $U$ from Table I  is $19.68$ which is very
far from zero. In our given model the maximally entangled ground
state obtained with \emph{U}=\emph{J}. This has a meaningful
physical interpretation which states that in such physical system
where the inter-atomic distance is very small, the on-site Columbic
repulsion \emph{U} can be very close to the inter-site Columbic
repulsion \emph{J}. Indeed the $H_2$ molecule is the best example of
such systems when the inter-atomic distance is minimum or
$d=0.7{\AA}$ and the $ab$ $initio$ value of U and V from Table II
gives $x=0.06$. Using $ab$ $initio$ parameters of Table I, we
obtained $x=0.06$ and this value of x gives maximally entangled
ground state. From Fig.2 the maximum entanglement value for $x=0.06$
is 1.99.\\ As we observe in Fig. 2, in the extreme limit of
$|\frac{U}{t}|\gg1$, i.e. strongly correlated systems the
entanglement becomes smaller and tends to 1. The non-magnetic
property of this state, sets a limit on minimum available
entanglement for this state. This point can be explained by using
eigenfunction listed in Table II. For $|x|\gg1$ one of the
$\alpha_+$ or $\alpha_-$ goes to zero and the other one becomes very
large. In both cases the ground state wavefunctions listed in Table
II reduce from extension on the four components to an extension on
the two components similar to $|E_2\rangle$ state. Therefore the
corresponding value of entanglement reduces from 2 to 1. This also
can be explained by tendency of the strongly correlated systems with
$|\frac{U}{t}|\gg1$ to unpaired electronic configuration in atomic
orbital such as $\varphi_2$ and $\varphi_5$ states in which energy
reduces by loosing \emph{U} term in Hamiltonian. Such states tend to
have parallel spin and
magnetism with a reduced entanglement.\\
In the case of $H_2$ where $t$, $U$ and $J$ are at the same order of
magnitude (See Table I) and $x$ is much close to zero hence the
molecule is in moderately correlated regime and one can obtain the
maximum available entanglement as it is shown in Fig. 2. In this
regime we have both spatial and spin correlated wave
function\cite{zar}. Neglecting exchange interaction in our model
($X_1=X_2=0$), significantly displaces energy levels of the system
(see Table II). However, the dependence of the ground and third
excited state wavefunctions on the exchange parameters only is
limited to $X_1$ and putting $X_1=0$ yields $x=0.07$ and similarly
the maximum entanglement becomes 1.99 which is the same as the case
of nonzero $X_1$ ($x=0.06$). In otherwords in order to obtain
maximally entangled states, the most effective parameters are direct
columbic interaction parameters $U$ and $J$ and exchange interaction
parameters do not alter the value of entanglement significantly.

\section{conclusion}
In conclusion, in this paper we applied a robust many electron
calculation on a simplest realistic two electron system i. e. $H_2$
molecule. Going beyond traditional Hubbard model let us to account
correlation effects accurately in the many electron wavefunction of
the ground and excited states. Using $ab$ $initio$ e-e interaction
parameters, indicates to a moderately correlated regime for the
molecule and resulted a maximally entangled ground and third excited
state. The wavefunctions of the not magnetic (S=0) ground and
excited states explicitly depend on correlation parameters whereas
the first excited states which is magnetic ($S^2=2$ and
$S_{z}\neq0$) is not entangled. The second excited state is not
magnetic but its wavefunction does not depend on correlation
parameters therefore it is a moderately entangled state. Anycase, by
switching on a magnetic field an entangled state with $S_{z}=0$ can
be extracted from a not entangled degenerate magnetic state. We
suggest that in a realistic molecular scale systems, there is two
criteria for finding maximally entangled electronic states, first
the system should be in moderately correlated regime and second the
system should have a non-magnetic $(S_{z}=0)$ electronic state.

\end{document}